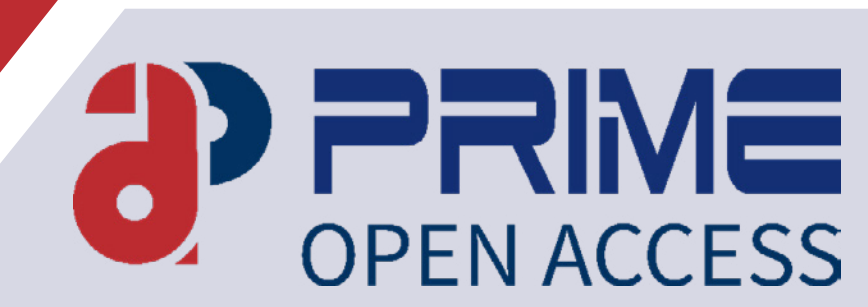

# Sensemaking in Sound from a Free-Response Final Exam for 6th Grade Physics

**Benjamin Ett[1,2*], Brice Le Roux[1] and Alice Delserieys[1]**

[1]ADEF Laboratory, INSPÉ, Aix-Marseille University, Marseille, France

[2]Massalia Potential Institute, Marseille, France

**Corresponding Author:**
Benjamin Ett, ADEF Laboratory, INSPÉ, Aix-Marseille University, Marseille, France.

**Citation:** Ett, B., Le Roux, B., Delserieys, A. (2026). Sensemaking in Sound from a Free-Response Final Exam for 6th Grade Physics. *J Theor Exp Appl Phys, 2*(3), 01-17.

**Declarations**
All authors contributed significantly to the article and approved the version to be published.

B.E. designed, collected, and analyzed the data and wrote the main manuscript. B.L. and A.D. helped shape the manuscript focus and main topics and reviewed the manuscript.

The authors declare that no funding was received to assist with the preparation of this manuscript.

The authors have no financial or non-financial competing interests to declare that are relevant to the content of this article.

The datasets analyzed during the current study are available from the corresponding author upon reasonable request.

## Abstract
This article investigates the sensemaking demonstrated in 6th grade students written responses to a single question free-response final exam for physics asking them to recount everything they learned over the course of the academic year. International exams such as the PISA and TIMSS show that students continue to have persistent difficulties with their comprehension, and appreciation, of science. Sensemaking and development of a deeper understanding of concepts is of fundamental importance when teaching science, but little progress has been made over the ensuing decades. The research questions are: What sensemaking of physics is communicated through a single question free-response exam? In particular, what topics are preferred by students and how are the semiotic resources of written and visual representations utilized to express sensemaking? With specific attention given to the topic of sound we observe two levels of comprehension, defined as Basic and Advanced, related to Bloom's Taxonomy, and we see evidence that previously low-performing students are capable of sensemaking at the Advanced level, thereby lending support to recent research calling for an increase in the level of complexity employed in primary and lower secondary science curricula. In addition to the sensemaking analysis, we discuss how these results are facilitated by the single-question free-response format which has been completely unexplored in the research literature and has the potential to be a valuable asset for research on or using assessment, as well as for teacher self-assessment.

**Keywords:** Affectivity, Assessment, Free-Response, Lower Secondary, Middle School, Physics Education Research, Sensemaking, Sound, Visual Representations

## Introduction

In this article we analyze the student responses on their final exam after a year-long, one hour per week physics class at the level of 6th grade (France 6ème, UK 7th). For the free response exam, students were given the following single question:

Discuss everything you learned in physics this year. Make sure to include formulas, definitions, drawings, and a discussion of the physical concepts. You will not be graded on your spelling or your grammar, only on your understanding of the material. (It is OK to put some words in your first language if you cannot remember the word in English.)

The students were not forewarned that they would have a test in this format and were not previously tested in this way, although they did have experience in answering open-ended questions posed by the teacher during the lessons. This free-response format allows us a unique view into which aspects of the course, including topics and explanations, were privileged by students in their responses. We will analyze these responses through the lens of sensemaking, in terms of how the affective teaching manner was represented in the exam responses as well as the advantages presented by the free-response format. This format has been historically under-utilized as a research tool in probing the nuances in students' comprehension of science, and its potential use in teachers' self-assessment has also gone unexplored. Through this single question, we will then show that the level of sensemaking attained across students of differing academic performance provides evidence that students are capable of comprehending abstract concepts of physics at a level thought previously to be too complex for their age. This analysis will be presented against the backdrop of decades of research into student misconceptions and the recent shift into studying the environmental and emotional factors present in their learning.

The central question at the heart of Physics Education Research is why do a majority of students find it so difficult to learn the fundamental concepts of physics, and why do these issues seem so resistant to intervention? The ineffectiveness of teaching physics was summarized decades ago by Redish [1]. "During the past decade, data have built up that demonstrate that as physics teachers we fail to make an impact on the way a majority of our students think about the world" while making a call for action "...we must change the way we are teaching...Society has a great need not only for a few technically trained people, but for a large group of individuals who understand science [1]." While society most certainly still has the need, it does not appear as if much progress has been made in the ensuing decades. While many studies have investigated students' knowledge of physics, from the naive theories they construct before formal instruction, to the misconceptions they continue to hold after years of study, a bulk of the research has focused on the students' side, considering their mental framework and how it changes (or not) throughout the learning and instruction process. The hope was that in understanding exactly how students think and where they go amiss, researchers could identify and eliminate these issues, but that has not been as effective as initially envisaged. Instead, when international assessments such as the PISA (Program for International Student Assessment) and TIMSS (Trends in Mathematics and Science Study) are considered, we see that there has been essentially very little to no improvement in the scores over the course of multiple decades for many countries [2,3]. This is why the quotes from Redish above still remain unfortunately very relevant. Being aware of the problem has yet to improve results. With this in mind, more recent research has focused on factors outside student responsibility, such as the teaching, curriculum, and environmental conditions that can influence student learning and engagement [4-6]. Although the number of perspectives to analyze the issue of student understanding has increased, students still do not seem to be making sense of the things they learn in physics.

Sensemaking as defined by Odden & Russ is "a dynamic process of building an explanation in order to resolve a gap or inconsistency in knowledge [7]. These explanations are built in one's own words, through an iterative process of construction and critique. Cognitively, when students are sensemaking they are building and refining their mental models, and they draw from and connect up multiple different representations and external explanations as they do so." The role of sensemaking is fundamental when considering how many students continue to misunderstand and incorrectly answer basic conceptual questions, even after receiving copious amounts of instruction on the subject [8]. That students are not making sense of the material they learn has been known for some time; as expressed in Shulman & Ringstaff, "It has long been recognized that teachers tend to ignore or at least give too little attention to the essential structures of subject matter and instead concentrate needlessly on the content of the subject as constituted of long, somewhat disorganized lists of facts, technical terms, or algorithms for problem solution [9]. It is unfortunately atypical for an instructor to take the trouble needed to communicate, in the variety of ways necessary, the underlying structures of his or her discipline. These structures are important precisely because they act both as organizers and as simplifiers of the otherwise endlessly long list of things to know within a given field.", as well as those more recently exploring the topic [10]. One of the most important factors in whether a student is able to learn science is their interest in the subject. In addition to scientific knowledge, the PISA and TIMSS also measure student interest in science. These tests have consistently shown that around half of students (already a low percentage) are interested when entering lower secondary and that this interest has dropped drastically by the time they reach upper secondary [11]. Addressing this drop, as well as the already low level of interest when entering lower secondary school should be one of the primary goals of research in science education both in terms of educating the next generation of scientists, as well as for having a scientifically literate public [12,13].

This article addresses the following research question: what sensemaking of physics is communicated through a single question free-response exam? In particular, what topics are preferred by students and how are the semiotic resources of written and visual representations used to express sensemaking?

**Background and Theoretical Framework**

Due to the perceived difficulty of learning physics, much of the education research has focused to a large degree on the student side of the responsibility. The study of student misconceptions attempts to understand and categorize how students bring naive theories of their physical understanding of the world with them before formal instruction, and why they carry persistent difficulties with certain concepts throughout their academic career [14-20]. In order to understand how to help students learn physics, researchers and educators need to understand the naive theories or misconceptions they have as well as how strongly they adhere to them [14]. Some of the first major breakthroughs came from the theory of "ontological categories" from the "Framework theory" and the concept of "knowledge in pieces" developed by diSessa [15-17]. Student misconceptions are important because of how persistent they are, which is seen at all levels from middle school to university [18-20]. Recently the situation has grown more nuanced in viewing these seemingly separate frameworks as different aspects of the same theory that can emerge dynamically when facilitated by teachers [21,22]. There are periods of disequilibrium which indicate the importance of conflict in learning and fits nicely within the framework of advanced sensemaking and pedagogies of discomfort. This idea of the emergent nature of an ontological category shift is viewed as a type of entropy in Volfson et al., which also concludes that educators should focus on teaching the underlying causal structure of emergent phenomena in order to promote deeper understanding [23].

While it is certainly important to classify the types of persistent misconceptions that students can carry with them throughout the learning process, there is the underlying assumption that the material was presented and taught effectively, but this warrants a more critical perspective. This is the purpose of research into the affective dimensions present in the learning environment [4]. Along this vein, Scott et al. discuss the discursive interactions of science classrooms and how authoritative and dialogical approaches can work in concert to promote engagement [24]. Considerable differences in conceptual change with respect to alternative teaching methods were seen in Barman et al. and Okur & Artun in their studies analyzing different teaching techniques for the concept of sound for similarly aged students as our study [25,26]. In our case, the teacher adopted an affective approach that attempted to create connection with the material through discussion of advanced physical concepts. Recent research investigating the teaching of advanced concepts in physics including Einsteinian physics, the kinetic theory of gasses, atomic-molecular theory, and the concept of fields supports this approach [27-30]. There is more evidence for this approach when considering the situation from a neuroscience perspective, where it was shown that the neural representations of concepts in physics are different between experts with a deeper understanding of fundamental phenomena when compared with undergraduate students, and that these might change systematically over time with continued exposure to deeper concepts [31]. This type of understanding is invaluable in the classroom, especially one given from a dialogical perspective and encouraging of questioning. If it is true that representations of concepts in the brain change with deeper exposure, then it would be reasonable to introduce higher level concepts at earlier ages.

The background of student misconceptions and the flexibility inherent in the free response format exam set the stage to analyze what sensemaking the students were able to communicate through their responses to the single question final exam. This paper will analyze sensemaking along the avenues of topic frequency and scientific discussion as well as visual representations, with particular focus placed on sensemaking of sound.

**Topic Frequency and Scientific Discussion**

The initial analysis of sensemaking includes determining and quantifying which physics topics the students mentioned, whether further elaboration through discussion was attempted, and whether images were utilized in conveying this information. The dialogical or rhetorical style in which the students present the information, as well as the imagery employed, distinguishes the different levels of sensemaking that have taken place. This will be further elaborated upon in the Methods and Results sections.

**Visual Representations**

Visual representations can be divided into two categories: "depictive" and "descriptive," defined by Schnotz. A description represents a subject matter with the help of symbols...Mathematical expressions such as...F=ma in physics [32]. Descriptive representations…have no similarity with their referent. A depiction, on the contrary, is a spatial configuration… Pictures such as photographs, drawings, paintings and maps are depictive representations...Depictions do not describe, but rather show the characteristics of an object.

Both depictive and descriptive visual representations can be utilized in parallel to promote flexible thinking and especially to avoid ambiguities when learning or problem-solving. Each plays a significant role and it is ideal to be able to use both, along with oral and textual representations to maximize comprehension.

Through the free-response format, this exam provides an opportunity to investigate which topics and visual representations have been privileged by the students after instruction over the course of the academic year. As sensemaking naturally encompasses comprehension, the use of visual representations are one of the main tools a teacher employs when trying to explain a concept beyond the written and spoken word. From a multiple representations perspective in addition to dialogical or written modes of representation, the visual element is an essential method of communicating information [33,34]. The multiple representation principle in the Cognitive Theory of Multimedia Learning states "it is better to present an explanation in words and pictures than solely in words [35]." In general, as a measure of metarepresentational competence in the ability to use visual representations productively as well as critique, modify, and create new ones, their use can be essential for analyzing the level of sensemaking. Other studies have investigated the role of visual representations in understanding and meaning making in specific situations such as shadow formation at the level of preschool, and for the understanding of "heat" at the primary level [36,37]. At the university level, multiple representations show their importance in problem solving and how the approach aids critical thinking when being able to view problems from multiple angles that are applicable across a wide range of problems in physics [38]. Multiple representations are discussed within the interplay of sense-making and fluency-building for student understanding in chemistry that can be highly dependent upon students' initial domain knowledge such that it can be beneficial for those with high prior domain knowledge, and detrimental to those without [39].

With respect to mental imagery, Chambers states that their work is "a clear demonstration that [mental] images are meaningful representations i.e. descriptive information must accompany depictive information [40]." Kosslyn makes the situation more concrete by showing that mental representations do in fact share the same mechanisms as visual perception highlighting the importance of visual representations [41]. However, we see that caution must be exercised during the constructions of these depictions considering that incorrect or redundant information can in fact impede knowledge acquisition which can also lead to counterproductive situations due to cognitive overload, but that individual differences and prior knowledge also play fundamental roles when analyzing the perception and interpretation of visual representations [42-44].

**Sensemaking in Sound**

In order to assess the sensemaking present in the student responses in a qualitative manner, the particular topic of sound was chosen for two reasons; sound is a natural phenomenon that almost everyone has direct physical experience with and because it was one of the most frequently discussed topics that also allowed students to really communicate their understanding to such a degree that nuanced sensemaking could be analyzed. This is in opposition to the other two highly mentioned topics (Force/4 Forces and Velocity/Speed/Acceleration) which were more straightforward and standard in their presentation by the students. The topic of sound is very interesting because even though most people are familiar, or even comfortable with the concept, misconceptions and difficulties with its comprehension are present at all levels from primary to university and post university physics graduates enrolled in teacher programs [8,23,45-47]. While there have been studies investigating student comprehension of the physics of sound at the level of lower secondary, as far as we are aware none have done so within the free-response exam format [18,48,49]. Many were given specific instruction for longer periods of time and questioned directly on the concepts presented whereas for our study the students were given the freedom to discuss whatever they felt was important and the topic of sound was covered for less than four of the total twenty hours of in-class instruction.

Due to the lack of specific studies concerning conceptions of sound, Eshach and Schwartz looked specifically at the preconceptions of sound, as a continuation of the work on substance schema and misconceptions of other physical phenomena [14,49]. Their study consisted of ten students in 8th grade with no previous experience studying sound phenomena. The authors saw that students' preconceptions were indeed that sound could be viewed as a substance, however this materialistic view was not necessarily consistent, yet these naive materialistic conceptions can be a good starting point. Although not stated explicitly in terms of Pedagogical Content Knowledge and Content Knowledge, the authors also suggest the important role played by the teacher in navigating these delicate situations where materialistic thinking can be both helpful, and a hindrance when learning about the true nature of sound waves. They discuss teachers using more non-verbal and visual representations as well as using more affective styles of teaching by combining the explanations of physical phenomena with historical anecdotes as ways of deepening students' understanding and levels of sensemaking.

A recent paper by Ravanis et al. discusses young students' (age 5-6 years) conception of different aspects of sound and confirmed that a large percentage of students view sound as a distinct physical entity, in line with results found in Hrepic et al. [50,51]. Similar findings were seen in West & Wallin where their study gave approximately 10 hours of instruction on sound to students aged 10 to 14 years old and tested (without a time limit) using a combination of short answer response and yes/no questions following a short prompt [52]. Besides showing positive results for the students, they also highlight that although possible at all age levels, results are highly dependent on the ability and knowledge of the teacher, echoing results of Keller et al. in discussing the important role of teacher pedagogical knowledge on student results [53]. It should be restated that in our case less than 20% of total class time was spent on the topic of "Waves and Sound" (see Table 1), and the students did not have the benefit of a prompt from the exam question itself to help with the recall of information.

In the present study, in order to measure the level of sensemaking achieved, student responses were divided according to their understandings of specific topics into two categories which we define as the two levels of sensemaking: Basic and Advanced, related to and based upon the first three levels of Bloom's Taxonomy [54,55]. The Basic level encompasses multiple initial learning techniques such as memorization and restating of facts accompanied with visual elements. The Advanced level, however, incorporates more outside knowledge and discusses processes and connection amongst multiple topics in a confident communication style. Essentially, the Basic level concerns itself with the first superficial knowledge of a topic, whereas the Advanced level demonstrates deeper connection with the material and big picture understanding. We choose three representative students based on multiple evaluation criteria such as the number and depth of topics discussed, the number of visual representations used, and the complexity and fluidity of their discussion. In relation to Bloom's Taxonomy, the Basic level aligns with the first baseline level "Remembering" in the revised taxonomy. Whereas the Advanced level aligns with "Understanding," it also touches upon certain definitions given as examples of "Applying," being the third level in the revised taxonomy. One can easily see how as the expectations for sensemaking at the Basic and Advanced increase with age and grade level, eventually most levels of the taxonomy would be encompassed. However, for our 6th grade students, only the first two levels are appropriate. Bloom's taxonomy was originally intended as a general framework and it was felt that the different disciplines of study would need to create their own specific taxonomies. This has been attempted within physics, however it has only been done at the university level [56,57]. Future research could and should consider attempting to construct such a taxonomy at the level of lower secondary.

**Assessment Format**

The question of how best to assess students' knowledge has been a continual theme in research but with unfortunately mixed results to date. The question of how best to evaluate has a long history of debate on the advantages and disadvantages of using closed or open-ended questions for testing. Martinez gives a nice general overview of important considerations when comparing specifically multiple-choice (MC) and free response (FR) testing formats, noting that trade-offs will always be present [58]. It concludes with the suggestion that a mixture of formats "can capitalize on their respective positive features while limiting their liabilities." There are very few results in the literature discussing this, especially within a science (or specifically physics) context and at the level of lower secondary. In a study at the university level, Dufresne et al. confirm the issue with correct answers to MC questions being false indicators of student knowledge and understanding [59]. Similarly, in a pair of papers by Simkin and Kuechler, the authors provide another good synopsis of the MC versus constructed response (CR) debate in relation to student understanding in the computer science domain [60,61]. They find that one of the main issues hindering explorations of the difference between question types is the treatment of either MC or CR questions as homogenous entities. Yet even when accounting for this, they find difficulty in creating MC questions that can attain the same knowledge level of CR questions. While most likely a step in the right direction, it still leaves the overall situation ambiguous as to whether MC questions can truly compensate for the knowledge demonstrated by CR questions. Searching for a middle ground, a study by Liu et al. regarding students in lower secondary science studying the concept of energy, attempted to construct a hybrid between FR and CR questions [62]. They introduce Explanation Multiple Choice response questions where normal MC questions are enhanced with questions asking students to choose between a set of explanations (which incorporate common student misconceptions as distractors) to clarify why they chose a specific answer to a MC question. The authors confirm previous extensive research that students are much better at recognizing correct explanations as opposed to generating them.

One of the more direct and comprehensive analyses was presented in the somewhat recent paper, where the authors looked at the effectiveness of the popular "concept inventories" standardized assessments for introductory physics within the context of 3D learning that encompasses not just student content knowledge, but also deeper concepts like scientific practices and crosscutting among scientific disciplines [63]. Not surprisingly they find that the concept inventories, which were developed before the creation of 3D learning, are insufficient at analyzing concepts beyond content knowledge. Another drawback is that these standardized assessments were mainly created for and used at the university level. They are also almost exclusively multiple choice. The authors speculate what a test that assesses broader learning outcomes would look like, stating directly that they "may also necessarily include tasks that are not multiple choice."

**Context and Participants**

The final exams represent the end product of a year-long introduction to physics course for 6th grade students at a private international school in France. The class met once per week for one hour and was composed of 21 students aged 10 to 12 years old, approximately half of whom came from countries where English was not their first language (the exam instruction allowing use of their maternal language was employed to help the students feel more at ease with the exam format). Students were required to take at least two science classes and the syllabus was designed to prepare students for the British IGCSE program which they would take when entering upper secondary. The final exam represents approximately 20 total hours of in-class instruction, not considering testing or review sessions.

The class was taught by an American male (first author) with a PhD in black hole theoretical physics in his first year teaching lower secondary school. Aside from mostly general topics, there was no standardized curriculum to follow, and the teacher felt free to discuss any topics of interest to himself after covering the requisite basics. As the teacher had extensive experience teaching at the university level for non-technical students, he accentuated the lessons with many examples, both historical and using current events in pop culture to help students have a more tangible comprehension, thereby establishing a deeper connection with the material. Considering the syllabus of the following years, this course was approached by the teacher as a first introduction to discovering the beauty of physics. The teacher used an affective approach to teaching in recognizing the importance of emotions in the sensemaking and learning process [5]. Getting a "feel" for the physical processes is essential in order to establish deep understanding and connection as expressed in [64,65]. Considering that a one-hour per week course is not intensive, it was of utmost importance to the teacher to concentrate on arousing interest and appreciation of physics. Minimal emphasis was placed on technical rigor so that the students could focus their attention on the general ideas and concepts behind the fundamental principles in a more qualitative manner. This was facilitated by the conversational and dialogical approach taken by the teacher when delivering information, including active solicitation of questions and directly encouraging engagement. As such, the teacher attempted to transfer his enthusiasm when discussing topics that were of interest to him by using many examples that students had direct experience with, with the hope of transferring the excitement of understanding the physical world to young students studying it in an academic context, possibly for the first time.

| **1. Units & Measurement 1hr** | **4. Energy 3hrs** | **5. The Earth and Beyond 3hrs** |
|---|---|---|
| ***2. Temperature 1hr*** | 4.1 What is energy? | ~~5.1 How do we see?~~ |
| **3. Forces 3hrs** | ~~4.2 Energy from the Sun~~ | ~~5.2 Day and Night~~ |
| 3.1 Introduction to Forces | 4.3 Energy Types | ~~5.3 What causes Seasons?~~ |
| 3.2 Balanced Forces | 4.4 Energy Transfer | 5.4 Stars & Our Solar System |
| 3.3 Friction & Air Resistance | 4.5 Conservation of Energy | ~~**6. Writing up of an experiment**~~ |
| 3.4 Gravity | 4.6 Potential and Kinetic Energy | ***7. Velocity & Acceleration* 5hrs** |
| ~~3.5 Tension and Upthrust~~ | ~~4.7 Elastic Potential Energy~~ | ***8. Waves & Sound* 4 hours** |

**Table 1: Syllabus for 6th Grade Physics Given to the Teacher Including the Approximate Time Spent on Each Topic (for a Total of Approximately 20 in-class Instruction Hours). A Strikethrough Line Denotes the Topic was not Covered and Italic Text Denotes Topics Added by the Teacher**

Considering the 20 hours of in-class instruction, this averages to approximately one class period per topic in the syllabus. For the specific topic of "Waves & Sound" the 4 hours represents approximately 20% of the total amount of class time and took around a month to complete. In order to pursue a deeper level of understanding and sensemaking for the most fundamental topics, certain topics were avoided while others were introduced in order to create a more cohesive collection. Connections were continually stressed and consistently referred back to when introducing new topics, such as when relating a discussion on Energy to the previously studied Force and Temperature. The class had no textbook until March, whereas prior to that, exercises and diagrams from various other sources were used for classwork and homework assignments. Homework was assigned every week, and the assignments were designed to take between 30-45 minutes, and were graded solely on effort, i.e. if the student attempted every question, full credit was awarded. Aside from the final exam and the homework, the other graded aspects of the class were the first semester exam (one hour) and three quizzes (30-min each), which were presented in the traditional format of short answer responses to direct questions, presented in the same chronological order as covered in class. No closed format (e.g. multiple-choice, true/false, etc.) testing was used.

## Methodology

Given the nature of the completely open response format it was determined that the general topic of sensemaking would be partitioned in four different avenues of analysis: three quantitative and one qualitative. The students based their responses on the information provided to them over the course of the year, which was a combination of the teacher's own knowledge of the subject, a previous year's textbook, and occasional resources from the internet.

The first and second avenues of analysis were the frequency of topics used by the students and whether an elaboration through scientific discussion was present in their responses. We define two categories, "mention" and "discussion": "mention" indicates that the specific word or topic appears somewhere in the student's exam, whereas "discussion" indicates an attempt to elaborate on the topic or concept. For example, an exam that stated "we studied forces" would count as the topic of Force being "mentioned," whereas if they went on to attempt a definition or explanation, that would be counted as "discussion" for the topic of Force. The specific topics are combined into a grouping of the nine most fundamental presented in class; listed chronologically these were Measurement/Units/Temperature, Force/4 Forces, Friction/Balanced, Gravity/Mass/Weight, Energy/Types/Conservation, Planets/Oumuamua/Comets, Velocity/Speed/Acceleration, Sound/Waves and the extra topic Black Holes/Cosmos. This allows us to see if there were any global topics in which multiple students had difficulty or a particular interest.

The third avenue of analysis concerns the visual representations utilized by the students in their final exam responses, which are analyzed by dividing all visual elements between the two categories of depictive representations (for images) along with descriptive representations (for mathematical formulas and information tables) in order to calculate simple averages. The quantitative approach will afford us the ability to discuss how sensemaking can be expressed through the visual dimension, and how the free-response exam can reveal subtle teaching and learning deficiencies that would not be as apparent on traditional style questions and exams.

The fourth avenue of analysis uses a more qualitative approach to discuss in detail the topic of sound, which was chosen because it was one of the most mentioned topics by the students, and because it has been a historically difficult topic to learn for students at all levels. For the sensemaking in sound analysis we can expand upon the two levels of sensemaking defined above within the specific context of the physical phenomenon of sound. Basic sensemaking includes definitions, formulas, and singular facts. There is an overall materialistic vision of sound where it is understood as a physical thing or object [15,21,22,49]. There is an understanding that sound is a wave that necessitates a medium in order to travel and that it has physical properties (e.g. wavelength, speed) that can be measured. This level is achieved in all three student examples. The Basic level can also include more technical aspects of sound such as understanding that there are different types of waves, where sound (longitudinal wave) is distinct from light (transverse wave), or that it comes from a source that vibrates.

To move from the Basic to the Advanced level, a student must realize that sound is not an object, but a process. They must go beyond simply listing all relevant properties and be able to coherently describe its behavior and how it is positioned within the wider context of physics. It should be understood that how a wave moves through space depends on the medium and its density, and that a wave can bounce off of surfaces (reflection or echo), move around them (diffraction), or change speed when traversing through them (refraction). It should also be understood that sound is a form of energy that propagates. Basic sensemaking makes the connection that sound travels, and how it travels, whereas sensemaking at the Advanced level concerns itself with the deeper notion of why it travels. After having defined two distinct levels of sensemaking, it was surprising to see that aspects of Advanced sensemaking can be present at all levels of student comprehension and prior academic performance. We demonstrate this explicitly with direct examples from all three students.

## Results and Analyses

### Topic Frequency

From Figure 1 we see that, in general and without the benefit of a question prompt, most of the students mentioned most of the topics. This demonstrates that most of the students possess a good general view of the topics that were presented throughout the year and were able to communicate them purely from memory. Even the least mentioned topic, Units, which was the first topic discussed in the course, was mentioned by over half of the students. Though very little class time was spent specifically on the topic of Units, the message of its importance within the grander scheme was apparently conveyed to a majority of the students because they chose to mention this topic specifically in their exam responses.

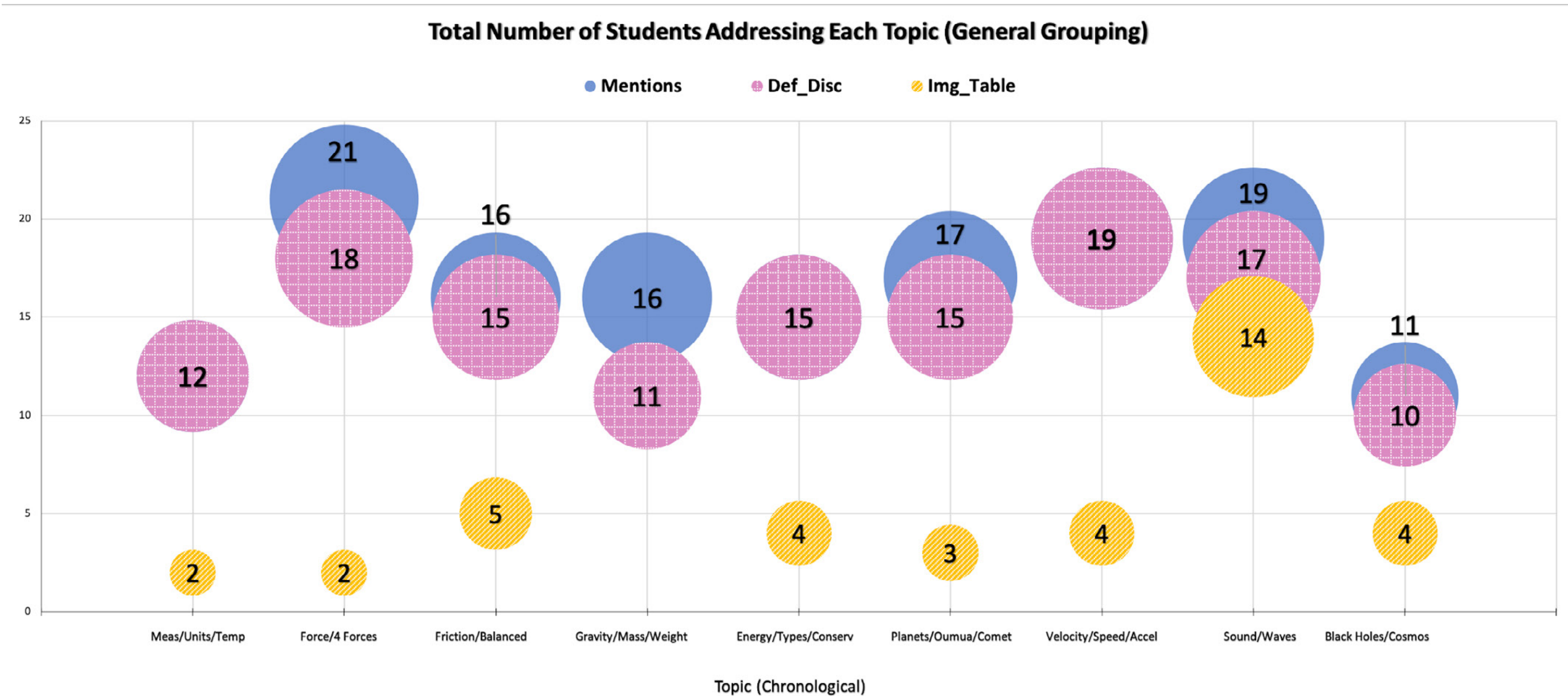

**Figure 1: Total Number of Students Addressing Each Topic with a Mention (Blue), Attempting a Definition (Pink) and Adding a Visual Representation (Yellow)**

We focus our attention on which topics were mentioned (blue circles), whether the student attempted to further discuss the topic (pink checkered circles) and finally, if they sought to employ an image or table (yellow circles) in their treatment of the topic. We can clearly see that at least 15 out of 21 students (over 70%) mentioned each of the seven general topics, excluding the first topic of Measurement/Units/Temperature and the extra topic of Black Holes/Cosmos (which themselves were still represented in over half of the student responses). We also see that most students attempted to further discuss or define each topic they mentioned, however apart from the topic of Sound/Waves, few students chose to include an image or table in their treatment. Overall this shows that a majority of the students were able to not only mention all of the most general topics presented over the course of the year, but that they also attempted to further define and discuss these topics. The extra topic of Black Holes/Cosmos is especially interesting due to the number of students that mentioned it considering that they were told specifically that it would not be on the exam, yet over half of the students found it important (or interesting) enough to include, possibly a result of affectivity [6,66].

**Visual Representations**

For graphing purposes, in Figure. 2 the depictive representations (solid color) occupy the bottom position, whereas the descriptive representations are further broken up into the two categories of formulas (dotted, middle position) and tables (stripes, on top). A total of 101 visual representations are displayed, 62 of which are descriptive (40 formulas, 22 tables), and the remaining 39 are depictive. This averages to approximately 5 visual representations per student, that constitutes 1 table, 2 formulas, and 2 images per student response. Only one student used no visual representations. 11 out of 21 students presented information in some form of table (5 of which used at least two), 17 out of 21 students used a depictive image (8 of which used 3 or more), and 19 out of 21 students used at least one formula (11 of which used two or more). An average of only 5 visual representations across 9 general topics implies the privileging of textual representation when communicating knowledge, far less than ideal. This is a key finding that would not necessarily be seen with more traditional type exam questions.

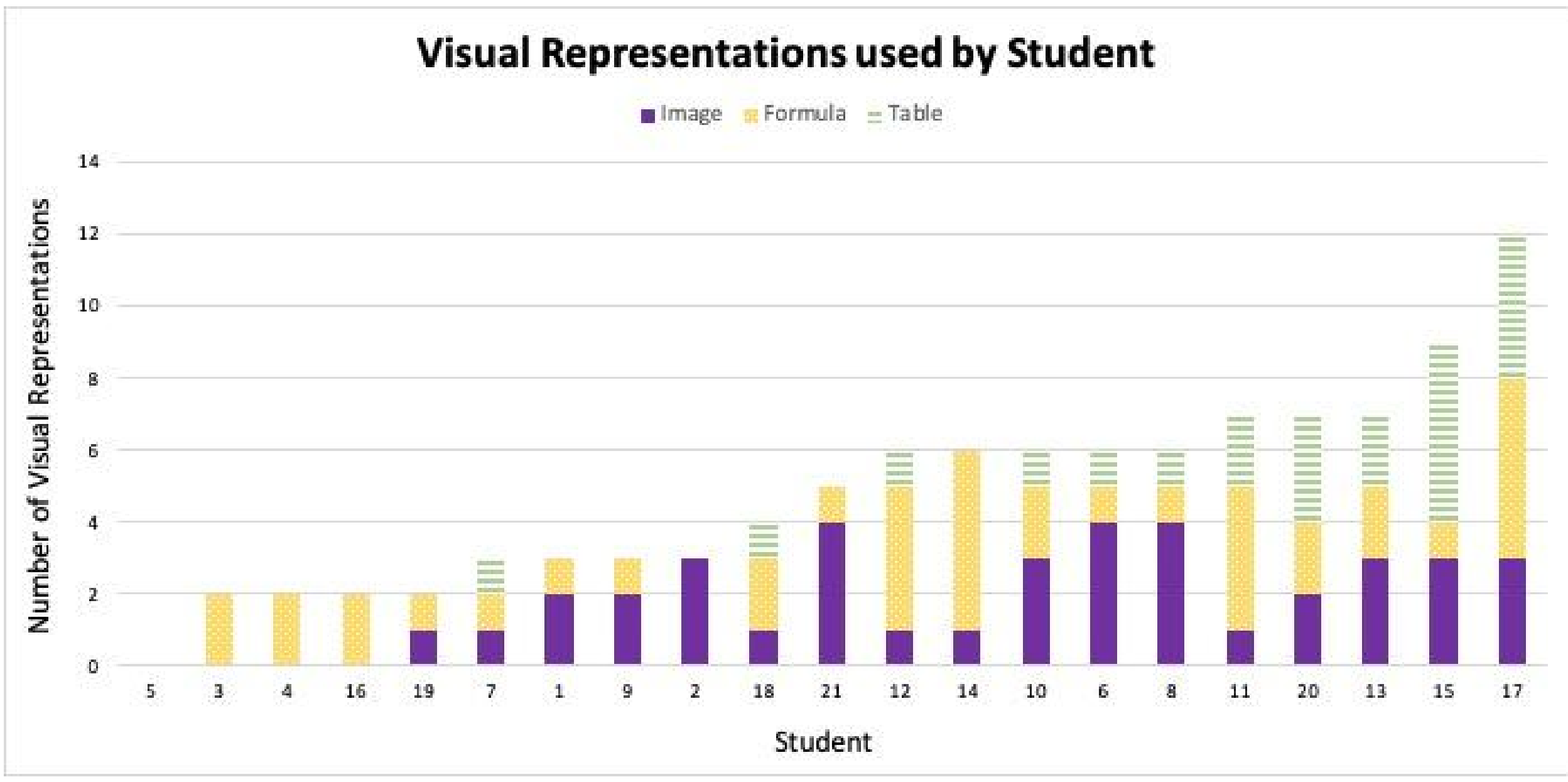

**Figure 2: Types and Number of Visual Representations Used by Each Student from Left to Right in Order of Least Total Representations used to Most**

For depictive visual representations, the topic of Sound/Waves is much more represented than the other topics. From the overall 39 depictive visual representations, 13 were in the topic of Sound/Waves, 11 of which being some type of waveform. Aside from this, the most visually represented (depictive) topics were those of Force (Friction), Force (Balanced), Planets/Space, and the extra topic of Black Holes, each with 4 depictive representations. These were the only categories with more than three depictive visual representations.

For the descriptive visual representation aspect, a different story emerges. The formulas for speed "s = distance/time" and for mass energy "E=mc2" account for 28 out of the 40 formulas present in the responses. Of the remaining 12, half were for acceleration and the rest were for various other topics. For the 22 descriptive visual representations of tables, the most used are for Temperature/Units with 5 representations, and the topics of Sound/Waves and Energy with 4 representations each. Only two students included no descriptive visual representation in the form of tables or formulas. 17 out of 21 students included some type of image or table, and 18 students included at least one formula, thereby adhering quite well to the exam instructions.

**Sensemaking in Sound with Students 2, 9, and 17**

**Student 2:** This student received one of the lower grades for the final exam, mentioning very few topics, and including minimal amounts of discussion. For the topic of sound, we see very little was demonstrated in terms of the Basic level of sensemaking, as essentially no definitions, formulas, or properties were directly discussed or listed. Although there is evidence of materialistic thinking where they state "sound goes in the hole" invoking the ontological category of objects, we also see a few indicators of the ontological category of processes and that Advanced sensemaking had taken place. Aside from the text, we can even see this through the visual representation of sound employed by the student, specifically that the initial action of "strumming" will eventually, after reflecting and resonating within the cavity, be emitted in the form of sound waves. There is a distinct four step temporal process to generate sound; step 1 – "strum", step 2 – sound goes "in the hole", step 3 – sound "hits all the sides", and step 4 – sound comes "out again." Steps 1 through 4 are displayed in a multimodal fashion, explained textually and shown visually, indicating an advanced understanding of the process underlying the behavior. Additionally, the sound process of a guitar was a homework question where the answer was discussed orally in class by the teacher, meaning this student remembered the picture from the book, as well as the description of the process, and was able to reproduce it textually and visually on the final exam without prompt. Even though this student's poor performance on the exam would indicate minimal or superficial understanding, the seeds of deep comprehension are present.

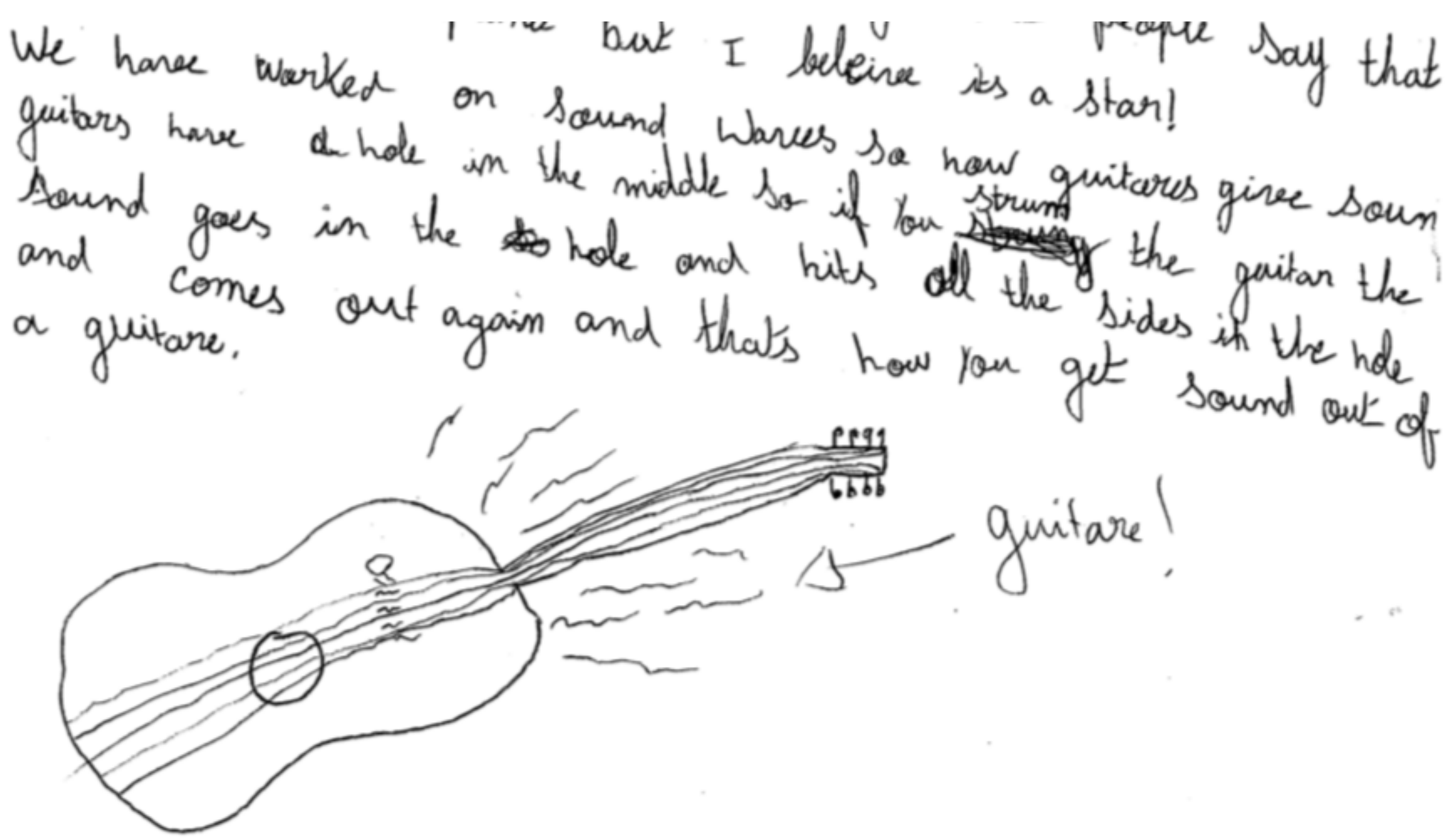

**Figure 3: Student 2 Visual Representation of Sound from a Guitar**

The specific example of a guitar was discussed in Volfson et al. where approximately one-third of the participants in the study held the idea of sound being a substance that is containable [46].

Employing the ontological categories for sound, the students show a materialistic view of sound such that it is a type of matter that can move from one place to another. This is an important first step into sensemaking on sound, however it is technically incorrect in the sense that sound is actually an emergent (as opposed to a direct) process. We would consider understanding this important technical fact as an example of Advanced sensemaking for older students, or those with previous experiences studying the topic. In our case, Student 2 is in 6th grade, whereas the students discussed in Volfson et al. are university and graduate level students concentrating in Engineering, Mathematics, and Physics that had previous experience learning about waves [46].

**Student 9:** This student was representative of those with a good understanding of certain concepts and a decent technical sophistication. For their exam, they chose to concentrate a majority of the time on the topic of sound. The student demonstrates sensemaking fully at the Basic level, along with traces of the Advanced level due to the use of multiple representations, communicated with coherent explanations and clear writing, thereby demonstrating comfortability with the topic.

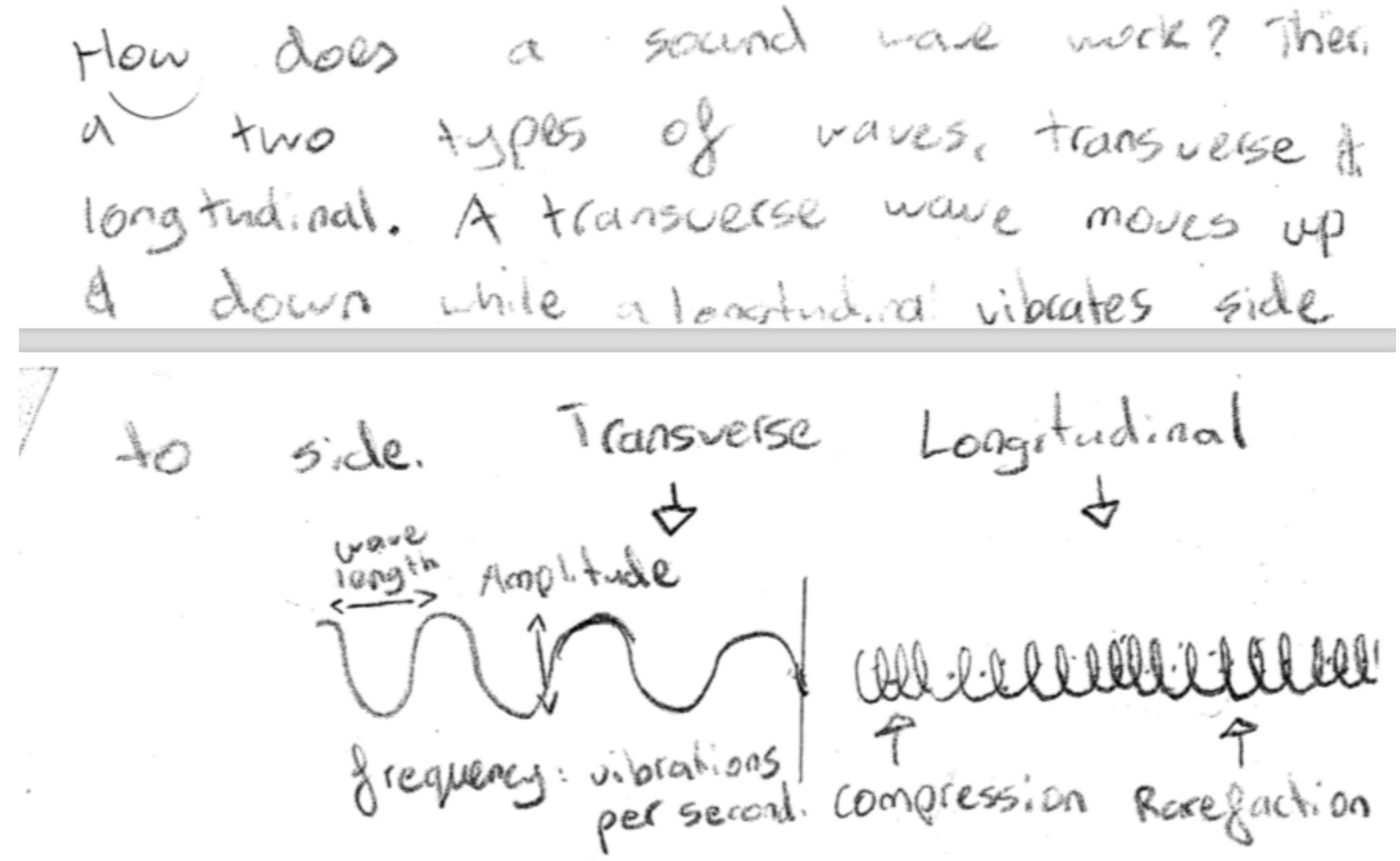

**Figure 4: Student 9 Combines Visual Representations Independently**

What stood out in this exam was that the student independently combined two separate visual representations of sound. The picture with the spring and particles were two different representations in the textbook, yet this student was able to successfully combine them and reproduce them on the exam without prompt. Furthermore, the combination of text and visual imagery shows multimodal learning (and communication) has taken place, and the student writes with a confident tone going beyond a mere listing of facts and formulas. In addition, the student is able to translate information tables viewed in the textbook into their own words, seamlessly integrating the contents into their presentation of the topic, which is impressive.

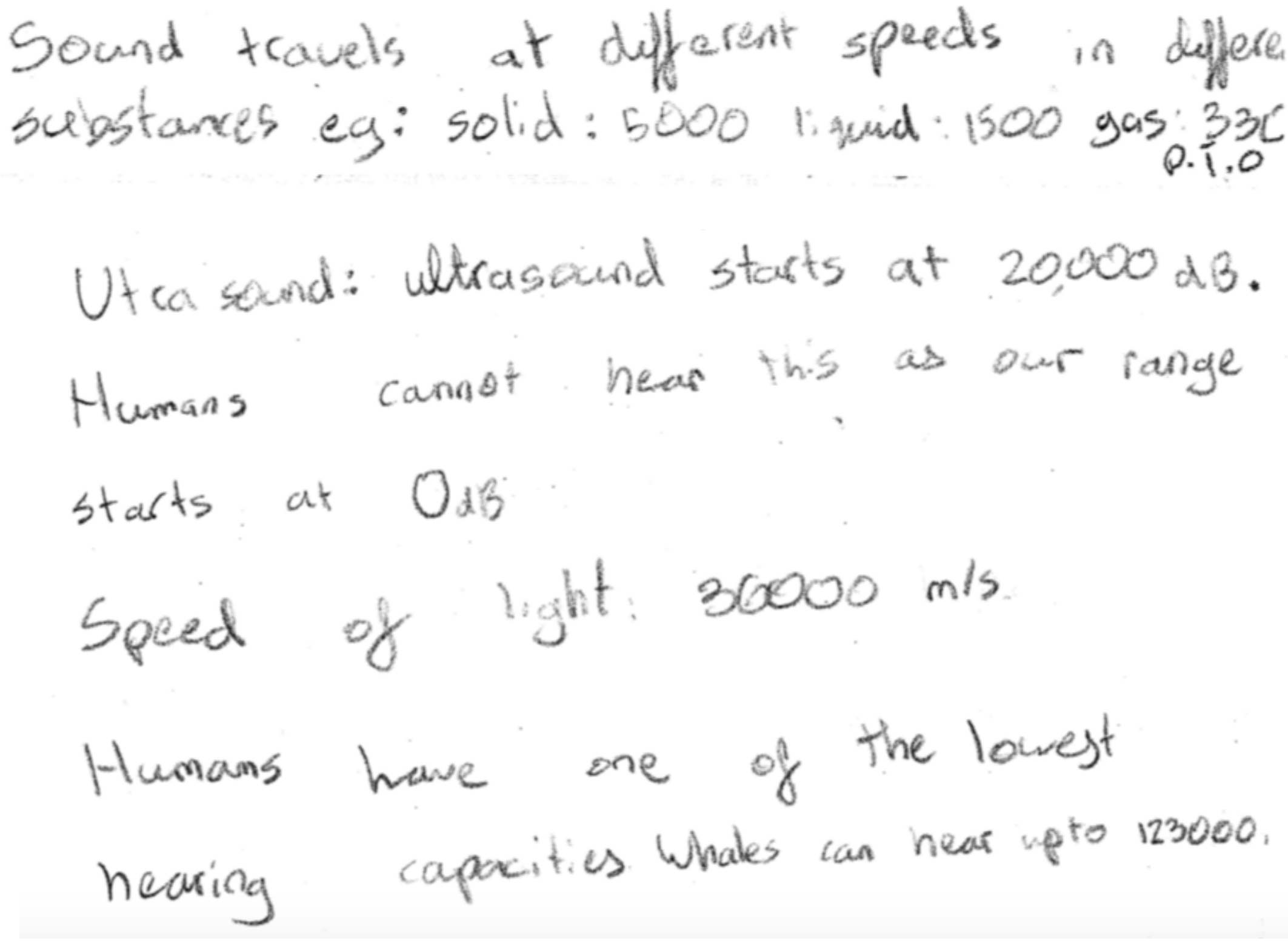

**Figure 5: Student 9 Reproduces Information Tables from the Textbook**

While some small mistakes are present, such as confusing Hertz 'Hz' for decibels 'dB', or an incorrect value for the speed of light, these are errors that are easily correctable. The student has demonstrated substantive conceptual understanding and would need only to refine the technical details. Excluding the specific formulas used for Sound/ Waves, they mentioned every other element listed in the requirements to achieve the Basic level of sensemaking. The breadth of the material discussed as well as the combination of depictive and descriptive visual elements, and the conversational tone demonstrated comfortability with the entire process and touches upon multiple features of sensemaking at the Advanced level.

**Student 17:** This student provided the most complete and thorough exam response. Beyond the topic of sound, they were able to reproduce almost every topic covered during the course of the year, as well as incorporating outside knowledge linking new information with the previous models of understanding. In addition, they used a fluid writing style that showed an advanced ability to communicate scientific understanding.

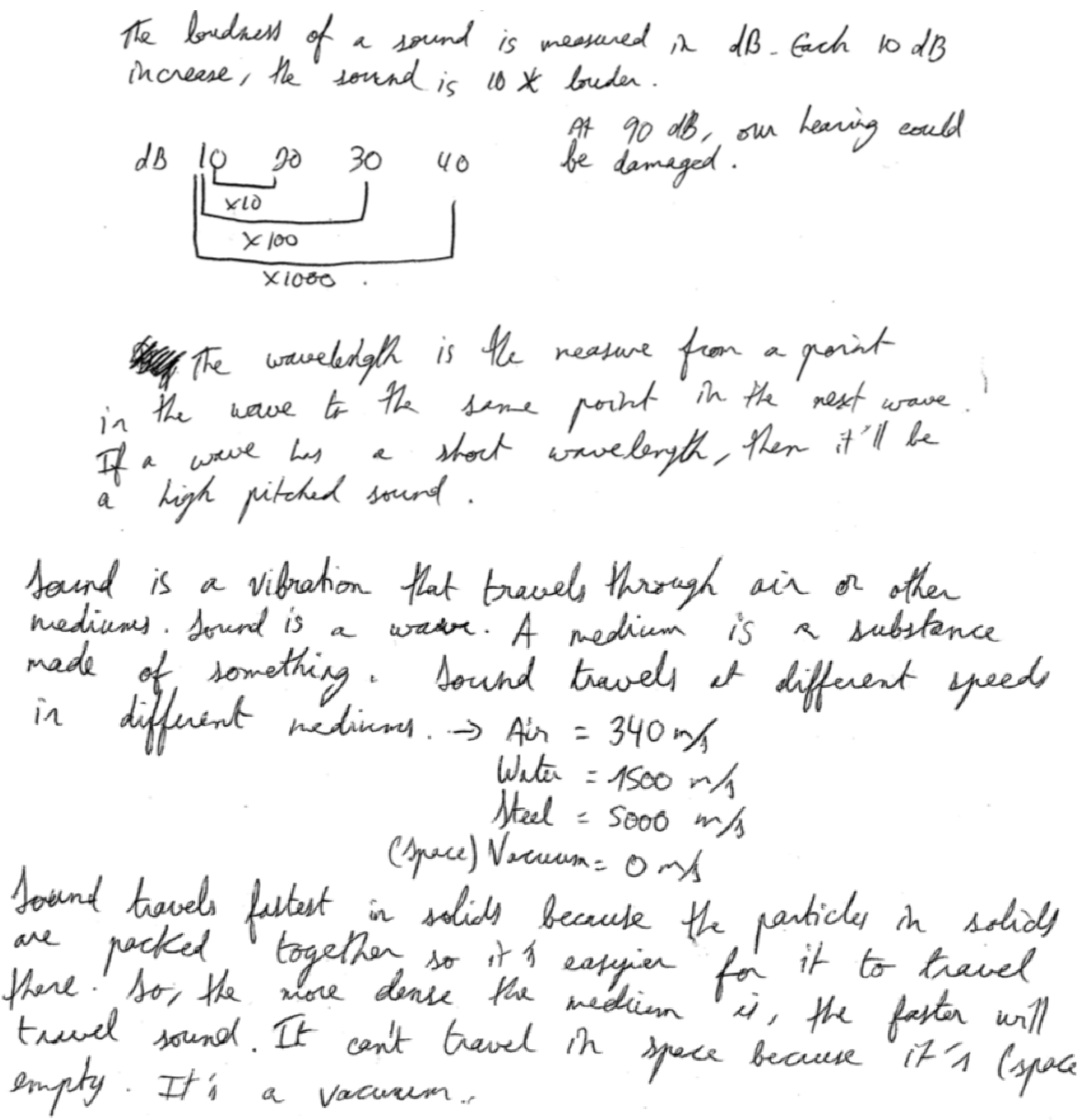

The loudness of a sound is measured in dB. Each 10 dB increase, the sound is 10 X louder.

dB 10 20 30 40

x10

x100

x1000.

At 90 dB, our hearing could be damaged.

The wavelength is the measure from a point in the wave to the same point in the next wave. If a wave has a short wavelength, then it'll be a high pitched sound.

Sound is a vibration that travels through air or other mediums. Sound is a wave. A medium is a substance made of something. Sound travels at different speeds in different mediums. → Air = 340 m/s

Water = 1500 m/s

Steel = 5000 m/s

(Space) Vacuum = 0 m/s

Sound travels fastest in solids because the particles in solids are packed together so it's easyier for it to travel there. So, the more dense the medium is, the faster will travel sound. It can't travel in space because it's (space empty. It's a vacuum.

**Figure 6: Student 17 Imports Learning and Reproduces Information Tables**

The visual depiction of the factor ten increase in the decibel scale was neither in the textbook, nor was it written by the teacher. This suggests that cognitive transfer may have taken place and was imported information from their math class. Aside from text, we see that both depictive and descriptive visual elements are employed. In addition to listing all relevant formulas and properties necessary for achieving the Basic level of sensemaking, a detailed discussion follows. The breadth and depth of the discussion, as well as the importation of outside material strongly suggests that Advanced sensemaking has taken place.

There are two types of waves:

Longitudinal wave: The particles move parallel to the wave.

Transverse wave: The particles move perpendicular to the wave.

Sound is a longitudinal wave while light is a transverse wave.

We talk by making our vocal cords vibrate. This vibratio projects energy into the air. The air in front of our mouth starts vibrating. The air molecules start moving back and forth from a place to another carrying the sound around. This movment is described as a wave We could use a slinky to represent a wave.

Longitudinal wave: Transverse waves:

compression rarefaction

There are 3 elements in a sound wave:

$\lambda$ = Wavelength (measured in meters) A = Amplitude (measured in meters)

F = Frequency (measured in Hz / # of waves/second).

$\lambda$

A

The amplitude is the measurement from the center of the wave, to it's highest or lowest peack. A sound with a large amplitude will have a high volume and a sound with a small amplitude will have a soft volume.

**Figure 7: Student 17 Reproduces, Textually, Explanations Heard in Class**

The part describing the process of talking and sound reception is especially interesting. The topic was discussed minimally in the textbook, yet was explained orally through physical and animated gestures by the teacher but not written on the board. This information was then reproduced in its entirety by the student. Furthermore and unique to this exam is the fact that sound is discussed as energy ("This vibration projects energy into the air. The air in front of our mouth starts vibrating,") which cements its classification in the Advanced level. This understanding moves beyond a materialistic view of sound to an understanding beyond those of older students, including those in university [18].

## Discussion

From what we have seen in our own results, these 6th grade students studying physics for the first time were able to, without prompt, recall and reproduce the fact that sound does not travel in space. This stands in contrast to multiple studies involving students at the upper secondary and even university levels. The immediate question is, is this because those more experienced students were never told directly that sound does not travel in space? Or maybe that it was not stressed by the teacher so it was quickly forgotten? Or perhaps it was stated multiple times, but not in a manner in which students were able to make connections? In the present study, the teacher not only specifically repeated the phrase multiple times, they also gave examples from popular culture, which were themselves reproduced on the exams. Interestingly, the anecdotes were only spoken by the teacher and not written down, meaning that the student that reproduced it on the exam either remembered it spontaneously during the exam, or decided it was important enough to write down in their notes when they heard it during the lesson. Considering the age and experience of the students, either possibility is significant.

The call for more explicitly teaching the concepts we want students to understand is echoed in Volfson et al. as a way to engage students to access deeper levels of understanding of physics principles, which we interpret in the context of our results as a path towards advanced sense making [46]. Considering the dynamic nature of sensemaking, what is advanced for a student in early secondary education, becomes basic as they move into secondary and post-secondary education. This new Advanced level would encompass a multi-faceted approach to knowledge in terms of breadth, depth, and importation of outside knowledge that is not possible during the early periods of education. Importation implies the cognitive transfer of outside knowledge has been utilized demonstrating connection and sensemaking, and is an example of filling in the gaps of knowledge. It is not surprising that regardless of the technical sophistication or surface level knowledge a student has about a particular topic, they are still capable of understanding some of the deeper and more fundamental aspects. Having this directly demonstrated in this fashion, is something that teachers can use to give themselves a detailed understanding of their students' knowledge not possible with solely MC or CR exams, and more readily accessible through FR formats. While it is true that FR questions can be time intensive to grade, perhaps more effort to incorporate these types of questions into exams can present a more well-rounded picture of student comprehension that is worthy of the increased time investment.

Students in high school and beyond can have two different approaches to how they study physics depending on their epistemological beliefs and whether they see physics as a loose collection of disconnected facts and formulas or as interrelated processes [66,68]. Students that spend too long trying to understand what is going on can become frustrated when their effort becomes counterproductive leading to demotivation and eventual disinterest. This is certainly the opposite of what most physics teachers would say they hope to hear. As stated in Gerace and Beatty, "If an instructor's exams ask students to recognize common problem types, manipulate equations, and calculate quantitative answers, then no reform of the course curriculum, methods, or assignments will convince them that they should be more concerned with conceptual understanding, qualitative and strategic analysis skills, and metacognitive awareness of learning [69,70]." This is where the free response questions are especially effective for analyzing broader and more nuanced student understanding. Through the single-question free-response exam we are able to analyze the learning environment in a more detailed manner, both in terms of student ability, affinity or miscomprehension, but also in how the teaching and presentation of the material could be improved. For our purposes, in addition to showing student content knowledge and misunderstanding or possible teaching improvements, the free-response format also allows us to analyze which topics, explanations, and aspects of the course were privileged by the students, thereby relating to questions of student affinity and identity, i.e. affective dimensions, with regards to physics. The lack of presence in the literature suggests that the free-response format is worthy of being considered more seriously as a teaching and research tool that can provide invaluable insights into the teaching and learning process at all levels of instruction. If the goal is advanced sensemaking, the long-form free-response format (as opposed to shorter responses to more direct prompts) can not only reveal the details in student thinking not necessarily seen by MC or CR question formats, but also be used as a tool for self-assessment by teachers. This approach can consider both the student misconceptions as well as dimensions of the teaching delivered in a more direct fashion than a carefully constructed multiple-choice test.

Within the context of the student responses it is important to discuss a more cautious interpretation of the results in terms of the images and tables used facilitated by the use of the free-response exam that could have gone unnoticed in more traditional type exams. While it's true that over half of the students used two images or tables, it seems that this number should be higher when considering that each student mentioned an average of seven topics. Even for the topic of Sound which had the highest use of depictive representations, it was only utilized by a little over half of the students (13 out of 21) with 11 of those 13 depicting a waveform. The topic of Sound also represented one-third of the total depictive visual representations (13 out of 39) used, showing that few students used visual representations for the other eight general topics taught in the course. One cannot say whether the students chose not to include more images or didn't think to include more images due to the time constraint or due to the surprise format of the exam itself. However this shows a bias towards textual representations, as opposed to visual, when students are asked to communicate their knowledge. This is an important realization, within the context of teacher self-assessment, into the privileging of textual over visual information within science, a domain where visual aids are taken to be crucial for understanding. Although visual aids were used for all topics, very few appeared on the final exam. This is a result that should be further explored in future studies on how teachers, already engaging in multi-modal teaching, can stress even further the importance of images for deeper learning [71-73].

In terms of reproducibility and suggestions, the affective nature of teaching as well as the utilization of the free-response format are two approaches to classroom intervention that are adaptable to most styles of teaching. The affective nature of teaching attempts to deviate from the traditional authoritative lecture format, incorporating the emotional side of the learning process. General guidelines are using a more dialogical presentation style, encouraging questions, using more tangible examples, stressing the fundamental concepts of the discipline, and making connections amongst everything that is taught. This allows students to better comprehend the discipline as a whole, instead of a collection of lists of facts to memorize. The free-response format also allows them to express any information they felt they were not able to show on the exam and a question in this form at the end of a traditional style exam could help to reduce test anxiety. While we are not claiming this present study to be definitive, the complete lack of literature discussing the possible benefits of using longform free-response questions warrants a more serious consideration not only in physics education research, but also possibly in science education research in general.

Limitations of this study are that the research potential of the exam responses were seen after the class had been given. Future research would incorporate student interviews, analyze the effect of stressing the importance of utilizing visual representations, and to develop a specific taxonomy for lower secondary physics related to Bloom's Taxonomy.

## Conclusion

Given the small amount of time spent in the classroom, it seems reasonable to assume that most sensemaking likely happens outside of class. While it is perhaps idyllic to imagine each student thoroughly engaged during a class discussion or exercise to the extent that the teacher can actively observe the sensemaking process, this is most likely unrealistic. In our study we saw that most students mentioned and discussed most of the topics that were covered in the class over the course of the academic year. We were able to distinguish two levels of sensemaking in their responses which we defined as Basic and Advanced, related to the first two levels of Bloom's Taxonomy. One of the more interesting aspects to emerge from the student responses is that we were able to see sensemaking at the Advanced level across all types of students, from low performing to high. This opens the possibility of discovering students with advanced understandings that could be underperforming in traditional settings as well as showing that younger students of all abilities are able to comprehend real concepts in physics thought to be too complex.

Every topic introduced by the teacher included a visual element and most of the students used some visual representation in terms of pictures, tables, or formulas, however it was across only a few topics and examples. Given the fundamental importance of visual elements to learning physics, it is a little disappointing that so few were employed in the students' exams. We cannot say whether they were actually thinking about visual representations and chose not to include them, or simply weren't considering them. This could reflect the teacher not sufficiently stressing their importance, or an outcome of the preference by students for textual over visual representations for explanation and discussion. Either way, this highlights the power of the single-question free-response exam in recognizing teaching deficiencies that could otherwise go missed and is an important avenue to explore in future research. The free-response format can be an invaluable tool in assessing the subtleties in student comprehension such that advanced levels of understanding were seen in students at all levels of academic performance. This format also allows the teacher to self-assess after finding systematic issues that would not necessarily be highlighted by exams in traditional formats. In the specific case of this study it allowed the teacher to realize that even though multiple images were presented with every physics topic introduced, they were not privileged by the students during the exam. Given the importance of mental representations, the teacher will be able to stress this aspect more in future situations. The free-response format should be more seriously considered for student evaluation in both research and academic settings. As stated by Redish, "To find out what our students really know we have to give them the opportunity to explain what they are thinking in words [1]. The information about the state of our students' knowledge is contained within them. If we want to know what they know, we not only have to ask them, we have to listen to them!" [1].